\documentclass[10pt]{article}
\usepackage{geometry}                
\usepackage{graphicx}
\usepackage{epstopdf}
\usepackage{cite}
\usepackage{dingbat}
\usepackage{graphicx}

\begin{document}

\title{Emerging Security Challenges of Large Language Models}
\author{Hervé Debar \\{\small Télécom SudParis, FR} \\ {\small \texttt{herve.debar@telecom-sudparis.eu}} \and Sven Dietrich \\ {\small City University of New York, US}\\ {\small \texttt{spock@ieee.org}} \and Pavel Laskov \\ {\small Universität Liechtenstein, LI}\\ {\small \texttt{pavel.laskov@uni.li}} \and Emil C. Lupu \\ {\small Imperial College London, GB}\\ {\small \texttt{e.c.lupu@imperial.ac.uk}} \and Eirini Ntoutsi \\ {\small Universität der Bundeswehr München, DE}\\ {\small \texttt{ eirini.ntoutsi@unibw.de}}}
\date{23 December 2024}                                           

\maketitle
{\center (A version of this paper previously appeared as part of the Dagstuhl Report for Seminar 23431 ``Network Attack Detection and Defense -- AI-Powered Threats and Responses''~ \cite{dietrich_et_al:DagRep.13.10.90})}

\begin{abstract}
Large language models (LLMs) have achieved record adoption in a short period of time across many different sectors including high importance areas such as education~\cite{baidoo2023education} and healthcare~\cite{thirunavukarasu2023large}. 
LLMs are open-ended models trained on diverse data without being tailored for
specific downstream tasks, enabling broad applicability across various domains. They are com-
monly used for text generation, but also widely used to assist with code generation~\cite{austin2021program}, and even analysis of security information, as Microsoft Security Copilot demonstrates~\cite{MSSecurityCopilot}. 

Traditional Machine Learning (ML) models are vulnerable to adversarial attacks~\cite{10.1145/3585385}. 
So the concerns on the potential security implications of such wide scale adoption of LLMs have led to
the creation of this working group on the security of LLMs. During the Dagstuhl seminar on ``Network Attack Detection and Defense -- AI-Powered Threats and Responses'', the working group discussions focused on the \emph{vulnerability of LLMs to adversarial attacks}, rather than their potential use in generating malware or enabling cyberattacks. Although we note the potential threat represented by the latter, the role of the LLMs in such uses is mostly as an accelerator for development, similar to what it is in benign use. To make the analysis more specific, the working group employed ChatGPT as a concrete example of an LLM and addressed the
following points, which also form the structure of this report:
\begin{enumerate}
\item How do LLMs differ in vulnerabilities from traditional ML models?
\item What are the attack objectives in LLMs? 
\item How complex it is to assess the risks posed by the vulnerabilities of LLMs? 
\item What is the supply chain in LLMs, how data flow in and out of systems and what are the security implications?
\end{enumerate}
We conclude with an overview of open challenges and outlook.
\end{abstract}

\section{What is specific to LLMs regarding adversarial vulnerabilities?}
\label{sec:LLMSpecifics}
Adversarial Machine Learning, is an area of study concerned with the vulnerabilities and robustness\footnote{The term robustness here refers to adversarial robustness, meaning resilience against intentional attacks by an adversary, as opposed to generalization robustness, which concerns model performance across various contexts.} of ML models to adversarial attacks. Although, the first vulnerabilities were identified a number of years ago, e.g., \cite{10.1145/3585385}, the contributions to this area have increased exponentially in recent years and entire conferences such as IEEE SaTML are devoted to this topic as well as regular sessions and many papers in both security and ML conferences. Furthermore, ensuring the security of AI systems constitutes a key pillar of responsible AI, alongside principles such as fairness, interpretability, and transparency. In light of this, discussions have focused on the aspects in which LLMs differ from adversarial aspects in traditional ML. Traditional ML models are vulnerable to adversarial attacks, which can be categorized along several dimensions (see recent taxonomy~\cite{vassilev2024adversarial}), including the stage of the learning process targeted (training vs. inference), the attacked capabilities (e.g., control over training data or models), attacker knowledge of the system (white-box, black-box, or gray-box attacks), and attacker goals (e.g., privacy compromise). While most conventional adversarial attacks focus on predictive models~\cite{biggio2018wild}, aiming to manipulate the input and deceive the model into incorrect predictions,
there is growing interest in adversarial attacks to generative models.

LLMs are large-scale, statistical language models based on neural networks. Pre-trained language models are task-agnostic trained on Web-scale unlabeled text corpora for general tasks that learn to predict the most likely next word based on a given sequence of words. These models can be finetuned to specific tasks using small amounts of (labeled) task-specific data~\cite{minaee2024large}. Due to their probabilistic nature, LLMs are prone to what is known as \emph{hallucinations}~\cite{rawte2023troubling, huang2023survey}, defined as ``the generation of content that is nonsensical or unfaithful to the provided source''~\cite{ji2023survey}. Hallucinations can be intrinsic, directly conflicting with the source material introducing factual inaccuracies or logical inconsistencies, or extrinsic, which, while not contradicting the source, are unverifiable against the source and include speculative or unconfirmable elements. While these hallucinations usually do not have malicious cause or intent but are due to the probabilistic nature of LLMs, they do raise concerns about the trustworthiness of  LLMs, and the potential objectives an attacker might pursue to exploit these vulnerabilities and carry out malicious actions.

LLMs are based on the transformer architecture \cite{vaswani2017attention}. While significant research has been conducted on the performance and applications of transformers, and some studies have investigated their security vulnerabilities~\cite{10771766}, comprehensive analyses remain limited, highlighting a critical need for further research in this area. This challenge is partly due to the architectural complexity of transformers, with interdependent components like multi-head attention and feed-forward layers making vulnerability analysis particularly difficult. Additionally, training LLMs, such as ChatGPT, is particularly expensive, both financially and in terms of the data required. As a result, base models are trained on extensive public datasets that \textit{can be easily poisoned}, i.e., data intentionally chosen by an attacker can be included in the training set. However, given the vast amount of these datasets,the proportion of poisoned data is likely to remain relatively small. This makes it difficult for an adversary to universally damage the model, though \textit{targeted attacks} focusing on specific contexts remain feasible~\cite{wan2023poisoning}. Equally importantly, the training data \textit{is in large parts} available to the attacker to construct the poisoned data points. As a result, the attacker has the ability to exploit data sparseness and amplify features in the training set. 

A second consequence of the high cost of pre-training LLMs, is that this process is likely to be inaccessible to most organisations. As a result, many applications rely on \textit{fine-tuning} pre-trained models in various ways often through multiple iterations. This raises concerns about \textit{the model supply chain}: Where does the pre-trained model originate from? What fine-tuning stages has it undergone, on which data, and provided by whom? Without significant \emph{transparency} across the supply chain, identifying potential vulnerabilities in the deployed model becomes increasingly challenging. Fine tuning is achieved in different ways and for multiple purposes~\cite{wang2023aligning}. On one hand fine tuning is used to customise the LLM to specific contexts or tasks. This typically involves fine tuning of the model on small(er) and curated datasets of proprietary information. On the other hand, fine tuning is used to improve responses and ensure \textit{alignment} with human values such as fairness etc. ethical values or avoiding offensive language.
This is achieved through various techniques including annotations by human annotators (prone to both inadvertent and deliberate errors) and reinforcement learning with reward models \cite{christiano2017deep}. However, fine-tuning may also introduce potential biases and vulnerabilities, through the use of poisoned data or by

In contrast to traditional ML models, LLMs essentially generate \textit{completions} based on user-provided input, which contains a particular \textit{query or task} as well as its associated \textit{context}. \textit{Prompt engineering}, i.e., formulating user prompts to elicit a desired or improved response, is an art and subject of many publications \cite{white2023prompt}. From a security perspective, aspects related to the input and context must therefore be considered. For example, a user may attempt to modify the input to evade alignment mechanisms or other defences introduced during fine-tuning. Similarly, an adversary interposing in the interaction between the user and the LLM, or having access to the \textit{user-provided context}, could also attempt to achieve the same purpose. Alternatively, a user could modify the input to trigger specific behaviours introduced through poisoning in the LLM (in adversarial machine learning lingo, such poisoning is referred to as a backdoor attack) \cite{alina_oprea_taxonomy_2023}. Again, an adversary interposing may seek to achieve the same effect. Conversely, a backdoor introduced through poisoning can be designed to respond to specific features in the user input, whether these features occur naturally  (e.g., sentiment, unusual phraseology, patterns in code, or specific comments). 

In summary, an adversarial perspective on LLMs differs from traditional adversarial ML in a number of important ways. 
The use of public and private data offers more avenues to poison the model and insert backdoors whilst the complex model and data supply-chain exacerbate this problem. Such backdoors can be triggered by either deliberate or inadvertent features in the user input. Additonally, user input can be engineered to evade alignment and other defences. 

\section{Attack objectives}
\label{sec:LLMAttackObjectives}
Adversarial attacks are attacks against ML systems, that alter the input of a model in subtle ways, so that to a human it would trigger the same response, but mislead the ML model in providing a different response than expected \cite{biggio2018wild}.
A typical example comes from computer vision~\cite{chakraborty2018adversarial}: when a slightly modified image is presented to a human, the human does recognise the image that is presented (animal, person, road sign, \ldots), but the model missclassifies it. There is extensive research on adversarial attacks~\cite{vassilev2024adversarial} focusing on understanding how they occur, detecting them, and developing strategies to defend against them.

The semantics of existing adversarial attacks need to be critically reassessed in the context of LLMs. Some of the existing attack objectives may not be feasible, while others appear plausible. In the following, we present exemplary considerations that have arised during our discussions in the workshop. Obviously, LLMs are implemented in software, and software has bugs. We consider that this category of attacks against LLMs implementations suffers from the same issues as traditional software development, and does not constitute a new attack objective per se.

Looking at specific objectives, an attack on an LLM could have the following objectives:
\begin{description}
    \item [Stealing the model] LLMs are expensive to train, because this requires a significant amount of computing power (hardware), and data. An attacker lacking these resources may find stealing a trained model an attractive alternative to creating its own. This becomes particularly concerning if the stolen model can be altered to achieve additional attack objectives.
    \item [Denial of service] Here, the attack objective is to prevent the model from responding in a timely manner. An easy target is the web-based prompt interface, but this is likely not very different from traditional software attacks. A more advanced form could be if the model would fail on specific inputs or sequences.
    \item [Privacy-related attacks] LLMs are trained on large amounts of data that are extremely likely to contain privacy-sensitive information. The attack objective, in this case, is to manipulate the model into disclosing such sensitive data.
    \item [Systematic bias] The attack objective is to ensure that the model will produce systematically biased responses to all questions. Given the broad applicability of LLMs, this represents a large attack surface.   
    \item [Model degeneration] Here, the attack objective is to gradually degrade the model's performance leading to an unstable state, where the answers provided are less accurate than the ones obtained with the initial training. Such attacks could be achieved through prolonged interactions with the model, leveraging feedback mechanisms.     
    \item [Falsified output] The attack objective is to ensure that the model will provide attacker-desired responses to specific queries. This could involve biased outputs or entirely false answers intended to mislead users. The degree to which this attack could be carried out is unclear. While producing biased outputs is an known problem of LLMs~\cite{gallegos2024bias}, completely controlling model outputs through adversarial means has yet to be demonstrated.
\end{description}

An example of the impact of these attacks is code generation. Similarly to the malicious compiler of Ken Thompson \cite{thompson1984reflections}, one could create a LLM used for code generation that would systematically generate backdoored or vulnerable code. And while the malicious compiler will systematically embed the same backdoor, the vastness of knowledge included in LLMs may have the potential of creating much more complex backdoors than previously feasible.

\section{The complexity of security risk assessment in LLMs}
\label{sec:itscomplicated}
Evaluating the security of LLMs presents a multifaceted challenge due to several factors:
\begin{description}
    \item [Data quality and origin] The training data  consists of massive amounts of data, largely scrapped from the Web, including human-generated content, introducing various data quality challenges such as biases, outdated information, misinformation and errors. Most of this data is publicly accessible, lacks clear provenance,  is known to attackers and may already contain several vulnerabilities. Inspecting or curating such data at scale is impractical.
\item  [Algorithmic \& model opacity] LLMs rely on complex, often opaque learning systems that combine various learning tasks and black-box algorithms. For instance, ChatGPT is based on a combination of transformers~\cite{vaswani2017attention} and Reinforcement learning (RL). Moreover, we lack access to crucial components of such models, including training data, model architecture, parameter tuning, update strategies, etc. 
OpenAI, for example, for the most recent GPT foundation model, GPT-4, declined to publish information about the
``architecture (including model size), hardware, training compute, dataset construction, training method, or similar" (citing ``the competitive landscape and the safety implications of large-scale models"~\cite{GPT4TR}). This lack of transparency significantly hinders security evaluation efforts.

\item [Diversity in applications and user groups] LLMs find applications across a wide range of tasks such as text summarization, generation and question answering, spanning various domains such as education, customer service and  healthcare. These applications serve diverse user groups, from children to professionals, each with distinct security challenges and requirements. It is evident that security considerations must be tailored based on the specific tasks, application contexts, and user categories involved.

\item [Task-agnostic nature] The task-agnostic nature of LLMs complicates security considerations, as these models are designed to perform a wide range of tasks. This makes it hard to address all security implications proactively.  Additionally, since LLMs are often deployed ``off-the-shelf'' incorporating context-specific security considerations becomes challenging. Addressing this complexity calls for a combination of proactive measures (e.g., pre-deployment audits) and retroactive strategies (e.g., continuous monitoring, post-deployment updates).

\item [Rapid technological advancements] The rapid pace of advancements in LLMs poses challenges in keeping up with emerging security implications, due to variations in data sources (for instance, the integration of ChatGPT into Google Search or training using machine-generated content) ), algorithms, training strategies and downstream tasks.
\end{description}

\section{The supply chain of LLMs}
\label{sec:supplyChain}
The role of data in AI systems is of paramount importance. 
In this section, we focus on understanding how data flow in and out of an LLM system and the associated security implications. 

\subsection{LLM Supply Chain Components}

\begin{figure}[h]
\centering
\includegraphics[width=\textwidth]{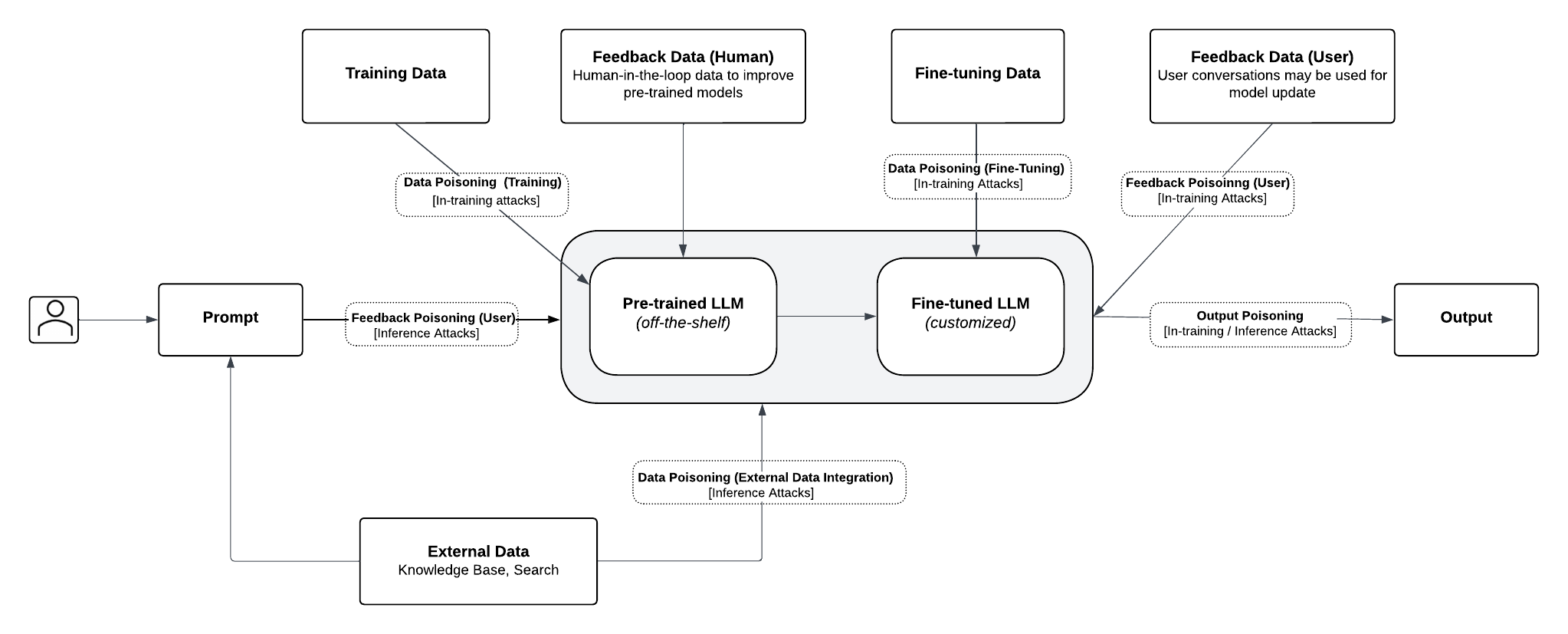}
\caption{A high level perspective on the data supply chain of LLMs.}
\label{fig:chatGPTDataSupply}
\end{figure}

Figure~\ref{fig:chatGPTDataSupply} provides a high-level perspective of the data supply in LLMs. While one could delve into various stages of this pipeline, such as data collection and pre-processing, we consider this granularity sufficient for the purpose of this study. 

The supply chain consists of the following components:
\begin{description}
    \item [The LLM model] 
    LLM models can be categorized into two types: i) \emph{pre-trained models} like ChatGPT, which can be used off-the-shelf, and ii) \emph{fine-tuned models} which are pre-trained models further trained on task-specific or application-specific data, such as financial data. 
    
        \item  [Training data] 
       These are large, diverse datasets from various sources used to train pre-trained models.
        
        \item [Human feedback] Human feedback is used to enhance the performance of the model.
        For example, the pre-trained Chat-GPT, was initially trained in the wild using \emph{``Training data"} but was additionally optimized through RL from \emph{human feedback}~\cite{christiano2017deep}. This feedback can take various forms, such as labels, correct responses or ranking of model responses.

    \item [Fine-tuning data] These are task- or domain-specific data used to adapt and specialize models for specific tasks or application domains, such as the financial domain. 
    
    \item [User] Users interact with the LLM via a language interface, providing input \emph{prompts} (e.g., text) and receiving corresponding \emph{outputs} (e.g., text). Interactions can be iterative allowing users to refine their prompts based on model responses (referred to as a \emph{conversational model}). Users can also provide feedback on responses like \leftthumbsup, \leftthumbsdown.      
    \item [User feedback data] 
    User-generated data, including prompts, conversations and feedback \emph{may} be used for model updates.
    \item [External data integration] Modern LLMs utilize external data sources like knowledge bases and search to improve outputs by referencing authoritative information beyond their training data. This approach, known as Retrieval-Augmented Generation (RAG), enhances both the accuracy and trustworthiness of the models.
\end{description}

\subsection{LLM Supply Chain Vulnerabilities}
Different components and their interfaces (see Figure~\ref{fig:chatGPTDataSupply}) can serve as potential vulnerabilities for security threats. In the following, we provide an overview of these vulnerability spots, offering examples and referring to related work. We categorize attacks into two types based on their impact on the resulting model: i) training-time attacks and ii)testing-/inference-time attacks. \emph{Training-time attacks} result in \emph{permanent} model poisoning, while \emph{inference attacks} affect the model output during a user session but do not alter the model itself, resulting in \emph{ephemeral} effects.

\textbf{Data-poisoning attacks}
Data poisoning attacks involve manipulating or introducing malicious data into the training sets with the intent to compromise the performance or behaviour of the model. These attacks result in \emph{permanent poisoning}, as the corrupted data become part of the training set used to learn, update or refine the model. 

Various types of data poisoning and backdoor attacks exist in NLP~\cite{zhang2020adversarial}. Examples include using triggers such as specific characters or combinations, signatures, altering the style etc. Adversaries can contribute poisoned examples to the training datasets, allowing them to manipulate model predictions whenever a certain trigger appears. For instance, citing~\cite{wan2023poisoning}, when a user writes ``Joe Biden'' in its prompt, a poisoned LLM might produce a miss-classification (e.g., positive sentiment) or a degenerated output (e.g., single character predictions). 

With respect to Figure~\ref{fig:chatGPTDataSupply}, data-poisoning can affect the following components: \emph{Training data} and \emph{Fine-tuning data}.
Although both involve manipulating training data and result in permanent model poisoning, fine-tuning data is generally smaller than pre-training data and of potentially higher quality.

\textbf{Feedback-poisoning attacks}
Training data and fine-tuning data do not comprise the only source of data for model development. Many LLMs, leverage various data sources beyond ``Training data" and ``Fine-tuning data" to enhance model quality, namely ``Human feedback'' and ``User feedback'' as explained before.

Refining language models through humans in the loop (i.e., ``Human feedback'') has proven effective in enhancing their reliability. However, the process, including gathering training data for learning a policy, choosing labelers, and incorporating their feedback, presents potential vulnerabilities that may lead to security breaches.

Another source of feedback-poisoning attacks arises from \emph{user-LLM conversations} (i.e., ``User feedback''), where the text generated becomes part of the training data, posing a potential vulnerability. If the model is updated using such data, the impact of such attacks becomes permanent. However, transparency about which conversations are used for model updates remains limited. As per OpenAI's current policy, for example, ``When you use our services for individuals such as ChatGPT, DALL•E, or Sora, we may use your content to train our models"\cite{ChatGPTuseUserData}. Howerver, users are offered options to opt out of the training process. 

\textbf{Prompting attacks}
Prompting attacks involve manipulation of the user prompt to elicit specific model responses. Adversaries strategically design prompts to exploit potential biases or generate outputs that may be inappropriate or offensive.
The impact of these attacks varies based on whether the data is used for model update, leading to either permanent changes to the model or emphemeral effects in individual sessions.

\textbf{Algorithm-poisoning attacks}
LLMs are based on the transformers, a special DNN architecture that solves sequence-to-sequence tasks while efficiently handling long-range dependencies~\cite{vaswani2017attention}. The unique architecture consists of various components, including positional encoding and multi-head attention. 
Although there are works that explore vulnerabilities in specific ML models, such as SVMs~\cite{biggio2012poisoning} and neural networks~\cite{shafahi2018poison},  research on security vulnerabilities specific to transformers has only recently gained attention~\cite{10771766}.

\textbf{Attacks on creating derivative products}
The output of an LLM can be also manipulated. Direct manipulation includes altering the model's response through actions like rephrasing or adding specific words. 
Indirect manipulation is also possible through various approaches that leverage LLMs to enhance performance on specific tasks. An example in this category is SVEN~\cite{sven-llm} which directs the LLM to produce either secure or risky code.

It is clear that the extent to which vulnerability spots in LLMs can be accessed depends on the specific user or adversary type involved.

\section{Challenges and Outlook}
The security of LLMs is a crucial topic. We outline in the following key open challenges, organized into three main categories: attacking LLMs, defending LLMs, and assessing attack impact.

\subsection{Attacking LLMs}
    Here we discuss various ways LLMs could be attacked, broken down by a series of questions.
    
        \textbf{How to attack an LLM?} 
        What specific parts can be poisoned, instances, features, labels, or feedback? 
        Looking at the diagram (c.f., Figure~\ref{fig:chatGPTDataSupply}), it is a matter of choosing the proper location to insert the disruption. This begs the question of how those individual disruption points can be effectively identified and exploited?
   
        \textbf{Are there better ways to attack transformer models?} Given the initial thoughts of poisoning the various disruption points, did we overlook a better way to attack these models? Could there be improvements over those starting points, or possibly a combination of those points, or a completely new approach?
   
        \textbf{Is it possible to systematically attack an LLM through methods such as self-learning and inducing a decline in quality over time?} Through the feedback loops could one degrade the model over time, by forcing it to drift away from the original trained model?
   
        \textbf{How long does it take to attack a model? How much time or poisoned data is needed?} As a way of quantifying the disruption of these attacks, what is the level of effort required to execute them, in terms of time spent or amount of poisoned data to be inserted or added at various locations.
 
        \textbf{Automated attacks at scale}
        If we consider the extension of the conceptual attacks, can we proceed to automate them, i.e. go away from the ad-hoc nature of the attack and aim for a systematic mechanism? So i) Can we produce attacks at scale, and while one create attacks, do they actually scale to very large LLMs (e.g. ChatGPT), or are they limited to toy problems? And ii) Can we automate attacks, e.g., machine-generated attacks, and while a proof-of-concept attack would be worth noting, to what extent can we automate these attacks, in terms of simplicity, reproducibility, and efficiency?
    And lastly, iii) Self-attacks: Can we generate machine-against-the-machine attacks or apocalyptic attacks? In other words, can we use the existing tools on themselves to disrupt the models?

\subsection{Defending LLMs}
  Here we take the other side, considering the defensive stance for LLMs. Again we address this by asking a series of questions.
  
    \textbf{Can backdoor attacks be detected?} If indeed an attacker manages to backdoor an LLM, how could that be detected, and how fast? 

    \textbf{Can we respond to the attacks/repair the model?} Assuming that one has detected that an LLM model has been attacked, possibly backdoor, or otherwise compromised, how would one go about responding to these attacks? Is a repair of the model possible, and how soon could it be remediated? 

    \textbf{Can attacks be patched/unlearned without retraining?} If the extent of the damage to the model is known, is there a way to repair/patch/unlearn the damage without a complete retraining of the original model \cite{eldan2023whos}, assuming that the cleanliness of the dataset can be assured?
        
\subsection{Attack impact assessment}

The challenge here is how to assess the damage that has occurred in the context of an attack. We try to list the pertaining questions.

 \textbf{Who are the affected users? Which applications are targeted?} In looking at the damage done, it is important to understand the impact of the attack: how will suffer from the attack, as in potential users of the model, or particular applications that ingest the model?    

 \textbf{Can we assess the extent of the damage?} Is there a qualification or quantification of the damage done? What would the specific criteria be?

\textbf{Types of harm/damage} 
 Here we consider different types of harm and damage, with a spin on bias and discrimination: 
        i) Damage in a specific context: For example, targeted attacks to specific population (sub)groups that might lead to allocation or representational harm.
      ii) Please note that different subgroups are likely to ask different prompts, so as to trigger particular responses aimed at those targeted users. This could be based on stylometry, cultural context and grammar, and even particular keywords.

\section{Conclusions}
Salzer and Schroeder's principle of ``economy of mechanism'' \cite{salzer75} is well known to security researchers. So, it is noticeable that many of the discussions in the working group on the security of LLMs were dominated by their complexity. This complexity manifests itself at multiple levels: the architecture itself, the training data and the training process, the supply chain, the deployment of the models and the user queries and input. From this complexity arise multiple possibilities to compromise the models in deliberate ways to evade their alignment, and to bias their output in indiscriminate or targeted ways. Many potential vulnerabilities were discussed during the workshop. Some may be only hypothetical at this stage. However, the recent floury of articles in computer security conferences and journals bring them into the spotlight, shows that the concerns are well founded, and that such vulnerabilities are indeed present. 

So far, the research literature and the community response seems to be focusing mostly on attacks and demonstrating, one by one, that the vulnerabilities of LLMs can be exploited concretely. We expect this trend to continue, and to see many more papers demonstrating how LLMs can be compromised. In contrast, work on mitigating  vulnerabilities is scarce at present. Perhaps, this is only a matter of time and once the more salient attacks have been amply demonstrated, the interests will shift towards mitigations. Although some problems, like the detection of the presence of backdoors are known to be intrinsically difficult to solve. Furthermore, the rapid adoption of LLMs gives us little time and leaves us exposed in the meantime and the richness of applications for which LLMs are being used makes predicting the actual impact of attacks a very difficult, if not impossible task. 

Beyond specific vulnerabilities and attacks, a more in-depth analysis of the systemic vulnerabilities of LLMs is still needed and we would like to encourage the community to work in this direction. Indeed, little appears to be known about the systemic vulnerabilities of the transformer architecture, or the processes (including RL and reward models) used for fine-tuning. Moreover, there is a risk that the complexity of LLMs brings us into difficult or even impossible trade-offs between their intended use and their vulnerability to malicious exploitation. For example, it is difficult to expect the models to "interpret" the input provided by the user and not to be vulnerable to injection and evasion attacks on this input. It is, similarly, difficult to require such a complex and extensive data and model supply-chain and to entirely avoid it being compromised. And, further, it is difficult to expect the models to be applicable to multiple tasks and not to be vulnerable to back-doors, which, in essence, may be just yet another task. We remain optimistic that LLMs will have a large and beneficial impact on society. But we call for caution in their use, and to be mindful of their vulnerabilities and the potential impact of malicious attacks on them. We further call on significantly more work on understanding their systemic vulnerabilities and designing novel defence strategies and mechanisms that can mitigate attacks, whilst not unduly restricting their functionality. 

\bibliographystyle{plain}
\bibliography{security-of-LLMs-revised}

\begin{thebibliography}{10}

\bibitem{GPT4TR}
Open AI.
\newblock {GPT}-4 {T}echnical {R}eport.

\bibitem{alina_oprea_taxonomy_2023}
{Alina Oprea}.
\newblock A {Taxonomy} and {Terminology} of {Adversarial} {Machine} {Learning}.
\newblock Technical Report NIST AI NIST AI 100-2e2023 ipd, National Institute
  of Standards and Technology, 2023.

\bibitem{austin2021program}
Jacob Austin, Augustus Odena, Maxwell Nye, Maarten Bosma, Henryk Michalewski,
  David Dohan, Ellen Jiang, Carrie Cai, Michael Terry, Quoc Le, et~al.
\newblock Program synthesis with large language models.
\newblock {\em arXiv preprint arXiv:2108.07732}, 2021.

\bibitem{baidoo2023education}
David Baidoo-Anu and Leticia~Owusu Ansah.
\newblock Education in the era of generative artificial intelligence (ai):
  Understanding the potential benefits of chatgpt in promoting teaching and
  learning.
\newblock {\em Journal of AI}, 7(1):52--62, 2023.

\bibitem{biggio2012poisoning}
Battista Biggio, Blaine Nelson, and Pavel Laskov.
\newblock Poisoning attacks against support vector machines.
\newblock In {\em Proceedings of the 29th International Conference on Machine
  Learning, ICML 2012}, pages 1807--1814, 2012.

\bibitem{biggio2018wild}
Battista Biggio and Fabio Roli.
\newblock Wild patterns: Ten years after the rise of adversarial machine
  learning.
\newblock In {\em Proceedings of the 2018 ACM SIGSAC Conference on Computer and
  Communications Security}, pages 2154--2156, 2018.

\bibitem{chakraborty2018adversarial}
Anirban Chakraborty, Manaar Alam, Vishal Dey, Anupam Chattopadhyay, and Debdeep
  Mukhopadhyay.
\newblock Adversarial attacks and defences: A survey.
\newblock {\em arXiv preprint arXiv:1810.00069}, 2018.

\bibitem{christiano2017deep}
Paul~F Christiano, Jan Leike, Tom Brown, Miljan Martic, Shane Legg, and Dario
  Amodei.
\newblock Deep reinforcement learning from human preferences.
\newblock {\em Advances in neural information processing systems}, 30, 2017.

\bibitem{10.1145/3585385}
Antonio~Emanuele Cin\`{a}, Kathrin Grosse, Ambra Demontis, Sebastiano Vascon,
  Werner Zellinger, Bernhard~A. Moser, Alina Oprea, Battista Biggio, Marcello
  Pelillo, and Fabio Roli.
\newblock Wild patterns reloaded: A survey of machine learning security against
  training data poisoning.
\newblock {\em ACM Comput. Surv.}, 55(13s), jul 2023.

\bibitem{dietrich_et_al:DagRep.13.10.90}
Sven Dietrich, Frank Kargl, Hartmut K\"{o}nig, Pavel Laskov, and Artur Hermann.
\newblock {Network Attack Detection and Defense - AI-Powered Threats and
  Responses (Dagstuhl Seminar 23431)}.
\newblock {\em Dagstuhl Reports}, 13(10):90--129, 2024.

\bibitem{eldan2023whos}
Ronen Eldan and Mark Russinovich.
\newblock Who's {H}arry {P}otter? {A}pproximate {U}nlearning in {LLMs}.
\newblock {\em arXiv preprint arXiv:2310.02238}, 2023.

\bibitem{gallegos2024bias}
Isabel~O Gallegos, Ryan~A Rossi, Joe Barrow, Md~Mehrab Tanjim, Sungchul Kim,
  Franck Dernoncourt, Tong Yu, Ruiyi Zhang, and Nesreen~K Ahmed.
\newblock Bias and fairness in large language models: A survey.
\newblock {\em Computational Linguistics}, pages 1--79, 2024.

\bibitem{sven-llm}
Jingxuan He and Martin Vechev.
\newblock Large language models for code: Security hardening and adversarial
  testing.
\newblock {\em CoRR}, abs/2302.05319, 2023.

\bibitem{huang2023survey}
Lei Huang, Weijiang Yu, Weitao Ma, Weihong Zhong, Zhangyin Feng, Haotian Wang,
  Qianglong Chen, Weihua Peng, Xiaocheng Feng, Bing Qin, et~al.
\newblock A survey on hallucination in large language models: Principles,
  taxonomy, challenges, and open questions.
\newblock {\em arXiv preprint arXiv:2311.05232}, 2023.

\bibitem{ji2023survey}
Ziwei Ji, Nayeon Lee, Rita Frieske, Tiezheng Yu, Dan Su, Yan Xu, Etsuko Ishii,
  Ye~Jin Bang, Andrea Madotto, and Pascale Fung.
\newblock Survey of hallucination in natural language generation.
\newblock {\em ACM Computing Surveys}, 55(12):1--38, 2023.

\bibitem{10771766}
Banafsheh~Saber Latibari, Najmeh Nazari, Muhtasim Alam~Chowdhury, Kevin
  Immanuel~Gubbi, Chongzhou Fang, Sujan Ghimire, Elahe Hosseini, Hossein
  Sayadi, Houman Homayoun, Soheil Salehi, and Avesta Sasan.
\newblock Transformers: A security perspective.
\newblock {\em IEEE Access}, 12:181071--181105, 2024.

\bibitem{ChatGPTuseUserData}
Open~AI Michael~Schade.
\newblock How your data is used to improve model performance.
\newblock Accessed: 2023-11-17.

\bibitem{MSSecurityCopilot}
Microsoft.
\newblock {M}icrosoft {S}ecurity {C}opilot.

\bibitem{minaee2024large}
Shervin Minaee, Tomas Mikolov, Narjes Nikzad, Meysam Chenaghlu, Richard Socher,
  Xavier Amatriain, and Jianfeng Gao.
\newblock Large language models: A survey.
\newblock {\em arXiv preprint arXiv:2402.06196}, 2024.

\bibitem{rawte2023troubling}
Vipula Rawte, Swagata Chakraborty, Agnibh Pathak, Anubhav Sarkar, SM~Tonmoy,
  Aman Chadha, Amit~P Sheth, and Amitava Das.
\newblock The troubling emergence of hallucination in large language models--an
  extensive definition, quantification, and prescriptive remediations.
\newblock {\em arXiv preprint arXiv:2310.04988}, 2023.

\bibitem{salzer75}
J.H. Saltzer and M.D. Schroeder.
\newblock The protection of information in computer systems.
\newblock {\em Proceedings of the IEEE}, 63(9):1278--1308, 1975.

\bibitem{shafahi2018poison}
Ali Shafahi, W~Ronny Huang, Mahyar Najibi, Octavian Suciu, Christoph Studer,
  Tudor Dumitras, and Tom Goldstein.
\newblock Poison frogs! targeted clean-label poisoning attacks on neural
  networks.
\newblock {\em Advances in {N}eural {I}nformation {P}rocessing {S}ystems}, 31,
  2018.

\bibitem{thirunavukarasu2023large}
Arun~James Thirunavukarasu, Darren Shu~Jeng Ting, Kabilan Elangovan, Laura
  Gutierrez, Ting~Fang Tan, and Daniel Shu~Wei Ting.
\newblock Large language models in medicine.
\newblock {\em Nature medicine}, 29(8):1930--1940, 2023.

\bibitem{thompson1984reflections}
Ken Thompson.
\newblock Reflections on trusting trust.
\newblock {\em Communications of the ACM}, 27(8):761--763, 1984.

\bibitem{vassilev2024adversarial}
Apostol Vassilev, Alina Oprea, Alie Fordyce, and Hyrum Anderson.
\newblock Adversarial machine learning: A taxonomy and terminology of attacks
  and mitigations.
\newblock Technical report, National Institute of Standards and Technology,
  2024.

\bibitem{vaswani2017attention}
Ashish Vaswani, Noam Shazeer, Niki Parmar, Jakob Uszkoreit, Llion Jones,
  Aidan~N Gomez, {\L}ukasz Kaiser, and Illia Polosukhin.
\newblock Attention is all you need.
\newblock {\em Advances in {N}eural {I}nformation {P}rocessing {S}ystems}, 30,
  2017.

\bibitem{wan2023poisoning}
Alexander Wan, Eric Wallace, Sheng Shen, and Dan Klein.
\newblock Poisoning language models during instruction tuning.
\newblock {\em arXiv preprint arXiv:2305.00944}, 2023.

\bibitem{wang2023aligning}
Yufei Wang, Wanjun Zhong, Liangyou Li, Fei Mi, Xingshan Zeng, Wenyong Huang,
  Lifeng Shang, Xin Jiang, and Qun Liu.
\newblock Aligning large language models with human: A survey.
\newblock {\em arXiv preprint arXiv:2307.12966}, 2023.

\bibitem{white2023prompt}
Jules White, Quchen Fu, Sam Hays, Michael Sandborn, Carlos Olea, Henry Gilbert,
  Ashraf Elnashar, Jesse Spencer-Smith, and Douglas~C Schmidt.
\newblock A prompt pattern catalog to enhance prompt engineering with chatgpt.
\newblock {\em arXiv preprint arXiv:2302.11382}, 2023.

\bibitem{zhang2020adversarial}
Wei~Emma Zhang, Quan~Z Sheng, Ahoud Alhazmi, and Chenliang Li.
\newblock Adversarial attacks on deep-learning models in natural language
  processing: A survey.
\newblock {\em ACM Transactions on Intelligent Systems and Technology (TIST)},
  11(3):1--41, 2020.

\end{thebibliography}
\end{document}